\documentstyle[aps,prl,manuscript]{revtex}
\gdef\journal#1, #2, #3, 1#4#5#6{               
    {\sl #1~}{\bf #2}, #3 (1#4#5#6)}          

\begin{document}
\draft

\title{Thermally activated reorientation of di-interstitial defects in 
silicon}

\author{Jeongnim Kim,$^1$ Florian~Kirchhoff,$^2$ Wilfried G.~Aulbur,$^1$ 
John~W.~Wilkins,$^1$ Furrukh~S.~Khan,$^2$ and Georg Kresse$^3$}
\address{$^1$ Department of Physics, Ohio State University, 43210}
\address{$^2$ Department of Electrical Engineering, Ohio State University, 
43210}
\address{$^3$ Institut f\"ur Theoretische Physik, Technische
Universit\"at Wien, Wiedner Haupstra\ss e 8-10/136, A-1040 Wien, Austria}
\maketitle

\begin{abstract}
We propose a di-interstitial model for the P6 center commonly observed
in ion implanted silicon. The di-interstitial structure and transition
paths between different defect orientations can explain the thermally
activated transition of the P6 center from low-temperature $C_{1h}$ to
room-temperature $D_{2d}$ symmetry.  The activation energy for the
defect reorientation determined by {\it ab initio} calculations is 0.5
eV in agreement with the experiment.  Our di-interstitial model
establishes a link between point defects and extended defects,
di-interstitials providing the nuclei for the growth.
\end{abstract}
\pacs{PACS numbers: 61.72.-y, 61.72.Cc, 61.72.Ji, 71.55.-i}
\newpage
Transient enhanced diffusion (TED) in boron-implanted silicon is the
limiting factor in controlling dopant profiles for submicron Si-based
devices. Interstitial defects in bulk Si generated during implantation
have been identified as the sources for boron
TED\cite{eaglesham94,stolk95,zhang95}.  A class of macroscopical
interstitial defects, namely $\{311\}$ defects, was suggested to emit
interstitials that can contribute to the enhancement of boron
diffusion under typical implantation
conditions\cite{eaglesham94,stolk95}.  Decreasing the ion implantation
energy can suppress the formation of macroscopic $\{311\}$ defects,
thus reducing boron TED.  However, boron TED exists even at very low
implantation energy (below 10 keV) in samples with no visible
$\{311\}$ defects\cite{zhang95}.  This implies that microscopic
interstitial defects, e.g., interstitial clusters, contribute to boron
TED.

The activation energy of boron TED, a measure of the energy required
to dissociate interstitials from the interstitial complexes, is lower
in samples without visible $\{311\}$ defects.  Interstitial clusters
can become important sources for boron TED at low temperature at which
the extended $\{311\}$ defects are still stable against dissociation.
Several defect states have been associated with interstitial
clusters\cite{lee76,lee98,benton97}.  One of them is the P6 center
commonly observed by electron paramagnetic resonance (EPR)
measurements in low-energy ion-implanted, proton- or
neutron-irradiated silicon\cite{lee76,lee98}.  The P6 center has
$\{100\}$ symmetry, distinct from the typical $\{110\}$ symmetry of
vacancy-related defects.  At low temperature (200 K), the symmetry of
the P6 center is either $C_2$ or $C_{1h}$ with a twofold axis parallel to
the $\langle 100\rangle$ direction.  A thermally-activated symmetry
transition occurs at room temperature (300 K) and the symmetry of the
P6 center becomes $D_{2d}$\cite{note0}. Motional averaging effects
have been suggested to cause the transition\cite{lee76,lee98}.
Complementary measurements of the $^{29}$Si hyperfine structure and
the stress alignment indicate that the P6 center arises from {\it
di-interstitial} defects.  The stable di-intersitital is an important
``precursor'' of interstitial clusters and more extended defects such
as $\{311\}$ defects. Yet, the structure and dynamics of the
di-interstitial have not been fully understood at an atomic scale and
no first-principle calculations are available to our knowledge.

In this letter, we propose a microscopic structure of a {\it
di-interstitial} whose low-temperature symmetry properties and
electronic structure are consistent with the P6 center. Furthermore,
the structure and transition paths between different defect
orientations (Fig.~1) can account for the experimental symmetry
transition to $D_{2d}$ at room temperature\cite{lee76,lee98}.  The
activation energy of the defect reorientation in our model is 0.5 eV 
in excellent agreement with 0.6 eV, the experimental activation 
energy extracted from the response of the P6 center to an external uniaxial
stress\cite{lee98}.  The experimental characterization of
the donor level is also consistent with
the calculated defect gap states.

The atomic and electronic structure of di-interstitial defects are
determined by {\it ab initio} total energy calculations within the
local density approximation (LDA) and generalized gradient
approximation(GGA)\cite{dftref,kresse}.  The local minimum and
metastable structures are obtained by structural minimizations using
the conjugate gradient method.  Starting geometries are stable
configurations obtained by molecular dynamics simulations and
structural relaxations using tight-binding
Hamiltonians\cite{kim97,kwon94,lenosky97}.  In the {\it ab initio}
calculations, the structural minimizations employ two supercell sizes:
(i) supercell A consisting of 120 atoms and (ii) supercell B
consisting of 72 atoms. The total energy and structure are well
converged with the plane-wave energy cutoff of 140 eV and 4 $k$-points
in supercell A\cite{conv}.

Figure 1(a) shows our di-interstitial model of $C_{1h}$ symmetry, with
the twofold axis parallel to the $z$ axis\cite{note0}.  The basic
constituents of the model are a center atom $I_0$ and dumbbell atoms
$I_1$-$I_2$.  Lee recently proposed a di-interstitial model composed
of the same building blocks\cite{lee98,leemodel}.  Our model differs
from Lee's model in (i) the orientation of the dumbbell atoms and (ii)
the location of the center atom.  The dumbbell structure, aligned
parallel to the $[110]$ direction, resembles the $\langle 110\rangle$
interstitialcy, the most stable point defect\cite{zhu96}.  Our
suggestion that the di-interstitial can be formed when an interstitial
is captured by the $\langle 110\rangle$ interstitialcy is supported by
the positive binding energy of the $C_{1h}$ di-interstitial with
respect to isolated interstitials (Table I).

Four equivalent $C_{1h}$ di-interstitial configurations, distinguished
by the location of the center atom, can be constructed with three
interstitials sharing one regular lattice site denoted as $o$ in
Fig.~2(a).  The {\it low temperature} symmetry containing a twofold
symmetry axis parallel to $\langle 100\rangle$ is consistent with the
$C_{1h}$ symmetry of our model\cite{lee76,lee98}.  Figure 2(b)
illustrates the four equivalent sites under $D_{2d}$ symmetry which
the center atom can occupy\cite{note0}.
Thermal averaging motions among four $C_{1h}$ configurations 
with a small activation energy barrier
can result in the {\it room temperature} $D_{2d}$ symmetry.

Indeed, we can identify transition paths that lead to the thermal
averaging motion between four local minima.  Three transitions --
denoted as $T_i$, $i = x,y,z$, according to the twofold symmetry axis
of its saddle point -- have the same energy barrier of 0.5 eV.

\noindent\underline{$T_z$ transition}: The transition
from the center atomic site $I_0$ to its mirror image $\bigotimes$,
{\it cf}, Fig.~1(a),
with respect to the $\{\bar110\}$ plane can occur by a displacement of
the atom $I_0$ along the $[110]$ direction.  The saddle point of the
$T_z$ transition has $C_{2v}$ symmetry with a symmetry axis along $z$.
The orientation of the dumbbell atoms remains the same after the $T_z$
transition.

\noindent\underline{$T_x$ transition} (Fig.~1(b)):
A displacement of $I_1$ along the $[101]$ direction results in a
$I_0$-$I_1$ dumbbell pair and the atom $I_2$ becomes the center atom
denoted as $I_0^*$.  The symmetry axis of the saddle point ($C_{2v}$
in Fig.~1(b)) is parallel to the $[100]$ direction.  The $T_y$
transition is similar to the $T_x$ transition, $I_2$ moving along the
$[011]$ direction.  As the result of the $T_x$ and $T_y$ transitions,
the orientation of the dumbbell atoms changes from the $[110]$
direction to the $[\bar110]$ direction.

The {\it room temperature} $D_{2d}$ symmetry can be explained by the
thermal averaging motion of three interstitials $I_0,I_1$ and $I_2$
alternatively occupying the four center atomic sites (Fig.~2(b)).
Note that the $T_i$ transitions leading to the thermal motional
averaging of the $D_{2d}$ symmetry involve a displacement of only one
of three interstitials along a $C_{2v}$ saddle point at each
transition.  The schematics of the potential surface along the high
symmetry path is presented in Fig.~3.  No other local minima are found
along the path $C_{1h}-C_{2v}-C_{1h}$.  The experimental $D_{2d}$
symmetry is very plausible within our model, since the low transition
energy barrier of 0.5 eV permits frequent $T_x,
T_y$ and $T_z$ transitions to motionally average the
four $C_{1h}$ configurations accessible to three interstitials.

\smallskip
\noindent {\it Structure and energetics of di-interstitial.}
Table I shows the formation energies $E_f$ and binding energies $E_b$ of the
$C_{1h}$ di-interstitial, the metastable $C_{2v}$ di-interstitial and
the $\langle 110\rangle$ interstitialcy. Our LDA
formation energy of the $\langle 110\rangle$ interstitialcy agrees
with that obtained by a previous LDA calculation\cite{zhu96}.  In
general, the bond lengths involved with the defect core atoms $I_0,
I_1$ and $I_2$ are slightly longer than the bulk bond length of
2.35\AA.  Charge density analyses confirm that the defect bonds are
indeed weaker than the ideal bulk bonds. The strongest bond for both
$C_{1h}$ and $C_{2v}$ di-interstitials is formed between the dumbbell
atoms $I_1$ and $I_2$.

Supercell B composed of 72 atoms is not large enough to quantitatively
describe the energetics and structural properties of the $C_{1h}$ and
$C_{2v}$ di-interstitials.  We find a reduction in the transition
energy barrier $\Delta$ from 0.6 (supercell B) to 0.5 eV (supercell A)
due to the relaxation of the third- and fourth-neighbor-shell atoms
from the defect core.  The bond lengths associated with the defect
core can differ by as much as 0.3 \AA\ between supercells A and B.
The structural properties of supercell B are {\it qualitatively
different} from those obtained by supercell A~\cite{cellB}.  However,
supercell A is sufficiently large for the converged energetics
and structural properties\cite{conv,tbref}.

We find large discrepancies between LDA and GGA in the formation $E_f$
and binding energies $E_b$, obtained by supercell calculations with
{\it different number of atoms}.  This is attributed to the larger
bulk cohesive energy of GGA compared to LDA.  On the other hand, the
relative stability of the $C_{1h}$ and $C_{2v}$ di-interstitials
indicated by $\Delta$ and structural properties are insensitive to the
choice of the exchange-correlation potential $V_{xc}$.  The formation
energy differences between supercells A and B are of the order of 0.1
eV within LDA. Similar differences are obtained with GGA.  The
dynamical process for the symmetry transition schematically shown in Fig.~3
is not influenced by $V_{xc}$. Typical differences between LDA and GGA results
are less than 0.005 \AA\ in bond lengths and 
less than 0.5$^\circ$ in bond angles for the same supercell size.

\noindent {\it Electronic structure of di-interstitial.}
The experimental analysis of the EPR signal of the positively charged
P6 center indicates that the donor level is strongly localized on the
center atom\cite{lee98}.  Figure~4 shows that the defect gap states of
the $C_{1h}$ di-interstitial are strongly localized.  The donor level
has strong $p$-character mostly localized on the atom $I_0$, while the
acceptor level is mainly localized on the dumbbell atoms $I_1$-$I_2$.
The donor level of the $C_{1h}$ di-interstitial is located at $E_v+0.1$ eV and
the acceptor level at $E_c-0.2$ eV.  

For the metastable $C_{2v}$ di-interstitial, the donor and acceptor
levels become almost degenerate and form a deep level at $E_v+0.4$ eV.
The metastable $C_{2v}$ di-interstitial becomes a stable defect when
it is positively charged.  The dashed line in Fig.~3 presents a
schematic potential surface along the high symmetry path.
Self-consistent calculations give a smaller energy difference of 0.14
eV between $C_{1h}^{++}$ and $C_{2v}^{++}$ configurations.
Our calculations on neutral and doubly-charged di-interstitials
further support the effective $D_{2d}$ symmetry of the
P6 center even at room temperature.

\noindent{\it Link to extended defects.}
The stable di-interstitial is the link between point defects and
extended defects. The dumbbell atoms of the di-interstitial are 
reminiscent of the $\langle 110\rangle$ interstitialcy, suggesting
the di-interstitial is formed by an existing $\langle 110\rangle$
interstitialcy capturing an interstitial.  Previously, we found a
stability hierarchy of intestitial defects from molecular dynamics
simulations with a tight-binding Hamiltonian\cite{kim97}: the
formation energy decreases in the order of interstitial clusters
$\rightarrow$ interstitial chains $\rightarrow$ $\{311\}$ defects. A
strong tendency of clustering of interstitials has been also predicted
from classical molecular dynamics simulations\cite{gilmer95}.  When
the interstitials are saturated such as in ion-implanted samples, the
di-interstitials can play the role of nuclei for the extended
interstitial defects, initially by providing interstitial sinks to
form elongated interstitial clusters and eventually interstitial
chains.  We note that di-interstitials, larger interstitial clusters,
and interstitial chains are composed of a common building block, the
$\langle 110\rangle$ interstitialcy\cite{kim97}.  The aggregation of
interstitial chains, in turn, can lead to the formation of extended
$\{311\}$ defects.

In conclusion, we present first-principle calculations for structural
and electronic properties of di-interstitial defects and relate our
stable di-interstitial structure to the P6 center.  The $C_{1h}$
di-interstitial model and the transition paths involving the
metastable $C_{2v}$ di-interstitial can account for different symmetries of
the P6 center both at low and room temperature.  The activation energy
of the defect reorientation is 0.5 eV in excellent agreement with
experiments. The localization of the donor level also agrees with
the experimental characterization.

We wish to thank Dr. Y. H. Lee for useful discussions.  This work is
supported by NSF and computational aids are provided by OSC, NCSA and
NPACI. FK acknowledges support by the NRL component of the DoD CHHSI
program.  The calculations have been performed using the ab-initio
total-energy and molecular-dynamics program VASP (Vienna ab-initio
simulation program) developed at the Institut f\"ur Theoretische
Physik of the Technische Universit\"at Wien.

\newpage
\begin{table}
\caption{Di-interstitial ({\it DI}\ ) formation energies $E_f$ and
binding energies $E_b$.  The binding energy is defined as $E_b = -(E_f
- 2E_f(I))$, where $E_f(I)$ is the formation energy of an isolated
$\langle 110\rangle$ {\it I}nterstitialcy.  Supercell A consists of 120
bulk atoms, while supercell B consists of 72 bulk atoms. The
orientation of the super cells is $[110]\times [\bar 110]\times
[001]$.  The energy barrier $\Delta$ is the total energy difference
between $C_{1h}$ and $C_{2v}$ di-interstitials for the same supercell.
}
\begin{tabular}{lccccc}
&& \multicolumn{2}{c}{Supercell A}
& \multicolumn{2}{c}{Supercell B}\\
&& LDA & GGA & LDA & GGA \\
\hline
$C_{1h}$ {\it DI}& $E_f$ & 4.93 & 6.01& 4.92 & 6.03 \\
&$E_b$ & 1.78 & 1.86 & 2.02 & 2.10 \\
\hline
$C_{2v}$ {\it DI}&$E_f$ & 5.40 & 6.51 & 5.48 & 6.60 \\
&$E_b$ & 1.30 & 1.36 & 1.46 & 1.53 \\
\hline
& $\Delta$ & 0.47 & 0.50 & 0.55 & 0.57 \\
\hline
$\langle110\rangle$ {\it I} &$E_f(I)$ & 3.35 & 3.93& 3.47 & 4.07 \\
\end{tabular}
\end{table}

\begin{figure}
\caption{(a) Atomic structure of a $C_{1h}$  di-interstitial
projected on the $\{\bar110\}$ and $\{110\}$ planes and (b) atomic
structures of the di-interstitial core during a transition between two
$C_{1h}$ configurations.  A regular lattice site located at the origin
of the axes is shared by three atoms -- (i) the center
atom $I_0$ and (ii) the $\langle 110\rangle$ dumbbell atoms $I_1$-$I_2$.
The site $\bigotimes$ is related to the site $I_0$ via a mirror
reflection across the $\{\bar110\}$ plane.  Three transition paths
$T_x, T_y$ and $T_z$ with the same energy barrier of 0.5 eV are
identified. The $C_{2v}$ symmetry of the saddle point has a twofold
axis parallel to the subscript of the transition $T_i, i=x,y,z$.  For
example, the transition $T_x$ in (b) displaces the center atom $I_0$
to $I_0^*$ site and the twofold axis of the saddle point in the middle
is parallel to the $[100]$ direction.  }
\label{f1}
\end{figure}

\begin{figure}
\caption{(a) Atomic structure of a $C_{1h}$ interstitial
and (b) schematics representing the four equivalent sites of the
center atom $I_0$ under $D_{2d}$ symmetry operations.  The regular
site indicated by $o$ is shared by three interstitials $I_0$, $I_1$
and $I_2$. The atomic indices are assigned based on the distance from
the origin and the symmetry to aid visualization.
The experimental $D_{2d}$ symmetry of the
di-interstitial at room temperature can be explained by the thermal
averaging motion of $I_0, I_1$ and $I_2$ alternatively occupying four
equivalent sites. As the result of a $T_i$ transition, the center atom
$I_0$ ``effectively'' moves to the corresponding site as indicated by the
arrows.  Note that black atoms lie below the grey atoms along $z$
axis.}
\label{f2}
\end{figure}

\begin{figure}
\caption{Schematic potential surfaces along the highly symmetric path of
$C_{1h}-C_{2v}-C_{1h}$ transitions for the neutral state (solid line)
and for the positively charged state (dashed lines) at a given
chemical potential $\mu$.  The neutral $C_{2v}$ di-interstitial is
metastable and the energy barrier $\Delta$ is 0.5 eV.  In
contrast, the positively charged $C_{2v}$ di-interstitial is stable
and the energy difference $\delta$ between $C_{1h}$ and $C_{2v}$
di-interstitials is only 0.14 eV.}
\label{f3}
\end{figure}

\begin{figure}
\caption{(a) Isosurfaces of gap state amplitudes of the $C_{1h}$ 
di-interstitial, (b) local density of states (LDOS) on the center atom
$I_0$ (solid line) and the dumbbell atoms $I_1$-$I_2$ (dashed line) of
the $C_{1h}$ di-interstitial and (c) those of the $C_{2v}$
di-interstitial.  The reference energy of (b) and (c) is $E_v$, the
valence band maximum of bulk silicon in the 120-atom supercell A.  The
calculated bulk conduction band minimum, $E_c$, is indicated by an
arrow and is about 0.8 eV from $E_v$.  The energy gap of the $C_{1h}$
di-interstitial is about 0.5 eV, while the donor and acceptor levels
of the $C_{2v}$ di-interstitial are almost degenerate.  LDOS of the
gap states is highlighted by the black (gray) areas for the donor
(acceptor) level.}
\end{figure}


\begin{thebibliography}{10}
\bibitem{note0}
The twofold axis of the $C_{1h}$ and $D_{2d}$ symmetry is chosen to be
parallel to $z = [001]$.  The $D_{2d}$ group contains two additional
twofold axes ($x$ and $y$) and $\{110\}$ and $\{\bar110\}$ reflection
planes.
\bibitem{conv}
We checked the convergence by increasing the number of k-points to 32
and the plane wave energy cutoff to 200 eV for the 72-atom supercell.
The formation energy of the $\langle 110\rangle$ interstitialcy
changes by 60 meV and the relaxed structure is indistinguishable from
calculations using 4 k-points and 140 eV. The energy difference
between the $C_{1h}$ and $C_{2v}$ di-interstitials is converged within
few meV. The structural properties and relative energetics are well
converged for supercell A calculations, confirmed by the less than 10
meV difference in the formation energies between 300- and 120-atom
supercell calculations using a TB Hamiltonian of Ref.~[12].
\bibitem{leemodel}
We find the di-interstitial model proposed by Lee~[5] is unstable. The
local minimum structure obtained by the conjugate gradient
minimization under $C_2$ symmetry constraints has a binding energy of
0.1 eV, much smaller than 1.8 eV of the $C_{1h}$ di-interstitial.
\bibitem{cellB}
The dumbbell atoms in supercell B are pushed toward the atoms $b$ and
$b^*$ making long bonds ($\sim 2.6$\AA) with the next-nearest-neighbor
shell atoms.
\end{thebibliography}
\end{document}